\documentclass{IEEEtran}
\usepackage{amssymb}
\usepackage{amsfonts}
\usepackage{amsmath}
\usepackage{algorithm}
\usepackage{graphicx,subfigure,cite,multicol,multirow,diagbox,booktabs,array}

\usepackage[dvips]{color}
\usepackage{float}
\ifCLASSINFOpdf
\else
\fi
\begin{document}
\title{Power Allocation Strategy of Maximizing Secrecy Rate  for Secure Directional Modulation Networks}

\author{Simin Wan, Feng Shu,  Jinhui Lu,  Guan Gui, Jun Wang, Guiyang
Xia, \\Yijin Zhang,  Jun Li, and Jiangzhou Wang

\thanks{This work was supported in part by the National Natural Science Foundation of China (Nos. 61771244, 61501238, 61702258, 61472190, and 61271230), in part by the Open Research Fund of National Key Laboratory of Electromagnetic Environment, China Research Institute of Radiowave Propagation (No. 201500013), in part by the Jiangsu Provincial Science Foundation under Project BK20150786, in part by the Specially Appointed Professor Program in Jiangsu Province, 2015, in part by the Fundamental Research Funds for the Central Universities under Grant 30916011205, and in part by the open research fund of National Mobile Communications Research Laboratory, Southeast University, China (Nos. 2017D04 and 2013D02).}
\thanks{~Siming Wan, Feng Shu, Jinhui Lu, Guiyang Xia, Yijin Zhang, and Jun Li are with School of Electronic and Optical Engineering, Nanjing University of Science and Technology, Nanjing, 210094, China.}
\thanks{~Guan Gui is with the College of Telecommunication and Information Engineering Nanjing University of Posts and Telecommunications, 66 Xinmofan Road, Gulou District,?Nanjing,?210003 China
Email: guiguan@njupt.edu.cn}
\thanks{Jun Wang and Feng Shu are also with  Fuzhou University, Fuzhou 350116, ~China.}
\thanks{Feng Shu is also with the College of Computer and Information Sciences, Fujian Agriculture and Forestry University, Fuzhou 350002, China, and the College of Physics and Information, Fuzhou University, Fuzhou 350116, ~China.}
\thanks{Jiangzhou Wang is with the School of Engineering and Digital Arts, University of Kent, Canterbury CT2 7NT, U.K. E-mail: j.z.wang@kent.ac.uk}


}
\maketitle

\begin{abstract}
In this paper, given the beamforming vector of confidential messages and artificial noise (AN) projection matrix and total power constraint,  a power allocation (PA) strategy of maximizing secrecy rate (Max-SR) is proposed for secure directional modulation (DM) networks. By the method of Lagrange multiplier, the analytic expression of the proposed PA  strategy  is derived. To confirm the benefit from the Max-SR-based PA strategy, we take  the null-space projection (NSP) beamforming scheme as an example and derive its closed-form expression of optimal PA strategy. From simulation results, we find the following facts: in the medium and high signal-to-noise-ratio (SNR) regions, compared with three typical PA parameters such $\beta=0.1, 0.5$, and $0.9$, the optimal PA shows a substantial SR performance gain with maximum gain percent up to more than $60\%$. Additionally,  as the PA factor increases from 0 to 1, the achievable SR increases accordingly in the low SNR region whereas it first increases and then decreases in the medium and high SNR regions, where the SR can be approximately viewed as a convex function of the PA factor.  Finally, as the number of antennas increases, the optimal PA factor becomes large and tends to one in the medium and high SNR region. In other words, the contribution of AN to SR can be trivial in such a situation.
\end{abstract}


\IEEEpeerreviewmaketitle

\section{Introduction}
Due to the broadcast nature of wireless transmission, security and privacy of confidential information increasingly becomes an extremely important problem in wireless networks. Directional modulation (DM), as a emerging and promising technique of physical layer security (PLS) in wireless  networks, has attracted tremendous research interests from both academia and industry world. The concept of secrecy capacity was proposed for a discrete memoryless wiretap channel in \cite{Wyner1975}, where the secure communication may be safeguarded if the channel of legitimated user is better than the channel of eavesdropper. Furthermore, artificial noise (AN) was utilized in \cite{Goel,Yang,zhao,Tao,Ding2014} to enhance the information-theoretic security. In two typical scenarios,  the transmitter with multiple antennas and the multiple-relay cooperation were used to improve the secret communication in \cite{Goel}. In \cite{Tao}, the authors proposed an AN-aided zero forcing synthesis approach for secure multi-beam DM, and the dynamic multi-beam DM was achieved by randomly changing the AN vector at the symbol rate. As such, the intended users could receive confidential information while illegitimated users could not successfully recover the confidential messages. Moreover, some symbol-level precoding and cooperative relays were employed in \cite{kalantari} and \cite{Lun,Hui,Zou2015} to enhance the PLS of wireless networks. Robust synthesis schemes for secure DM were proposed in \cite{Hu2016,Shu2016,Zhu,Xuling} to enhance the security performance of desired directions and distort the constellation points of undesired directions. Given the uniform distribution of direction of arrival (DOA) measurement errors, the authors can significantly improved the bit error rate (BER) performance  based on minimum mean square error criteria in \cite{Hu2016}. In addition, the authors of \cite{Shu2016,Zhu,Xuling} extended the idea of literature \cite{Hu2016} to multi-beam DM scenarios in broadcasting systems, multicast precoding and multi-user multiple-input multiple-output (MIMO) systems in the presence of direction angel estimation errors. Furthermore, the authors in \cite{shufeng2018} proposed a low-complexity secure and precise wireless transmission scheme combining random subcarrier selection (RSCS), orthogonal frequency division multiplexing (OFDM), and DM. In such a concept, the beamforming vector forms a two-dimensional direction and distance dependent property, which can transmit confidential messages to any given position, and form a high receive power peak around the position with only a little energy leaking out to the undesired area, where the undesired area is composed of all areas outside a small neighborhood around the desired position.

 In \cite{Tsai,Yongpeng,Xiangyun,TongXing,YAN,LinHu,Huanhuan}, the transmitter transmitted confidential information concurrently with AN and the optimal power allocation (OPA) was analysed in different scenarios. Lower bounds of secrecy rate (SR) in multiple-input  single-output with single-eavesdropper (MISOSE), multiple-input single-output with multiple-eavesdropper (MISOME) and multiple-input multiple-output multiple-eavesdropper (MIMOME) were derived in \cite{Tsai}, and the closed-form solutions of  OPA were obtained from these bounds. Additionally, equal power allocation (PA) and water-filling PA were analysed in this paper. The authors of \cite{Yongpeng} investigated the impact of PA parameter based on the asymptotic achievable SR in MIMO system with an active eavesdropper when the number of transmit antennas was infinite. Additionally, the effects of imperfect channel state information (CSI) were considered in \cite{Xiangyun,TongXing}. In \cite{Xiangyun}, the OPA for the noncolluding eavesdropper case and colluding eavesdropper case were discussed, respectively. The authors of \cite{TongXing} proposed the OPA strategy in the presence of spatially randomly distributed eavesdropper. Furthermore, the authors designed a correlation-based PA strategy and compared it with uniform PA and OPA in \cite{YAN}, where the transmitter was equipped with correlated antennas. A cooperative jamming scheme was proposed in \cite{LinHu} to enhance the PLS, and the authors analysed the impact of PA parameter between confidential information and AN by minimizing the secrecy outage probability subject to a minimum SR constraint. In \cite{Huanhuan}, the normal transmitter DT is responsible for broadcasting public information to its service subscriber DR and disrupting the unauthorized eavesdropper was taken into consideration to guarantee secure communication.

However, all the above literature concerning PA does not belong to the scope of DM.  To the best of our knowledge, there is still no research  investigation of PA in directional modulation networks. In this paper, given any beamforming vector and AN projection at DM transmitter, we propose an OPA strategy to maximize the SR. Simulation results confirm the benefit of the OPA.  Our main contributions are summarized as follows:
\begin{enumerate}
  \item Given the beamforming vector of confidential messages and AN projection matrix and total power constraint,  a PA strategy of maximizing secrecy rate (Max-SR) is proposed for secure DM networks. By the method of Lagrange multiplier, the analytic expression of the proposed PA  strategy  is derived.
 \item Take  the null-space projection (NSP) beamforming scheme as an example, its simple closed-form expression is also derived. From simulation results, it follows that the PA has an obvious dramatic impact on the SR performance. Compared with three typical values of PA factor such as $\beta=0.1, 0.5$, and $0.9$, especially with small-scale number of transmit antennas at  DM transmitter, the  SR performance gain achieved by the OPA is relatively attractive with the maximum SR improvement percent being  more than $60\%$.
\end{enumerate}

The remainder of this paper is organized as follows. Section II presents the DM system model. In Section III, the PA strategy for Max-SR is proposed and its closed-form expression is given. Subsequently, the NSP scheme is taken as a special example and its OPA expression is simplified. Simulation and numerical results are shown in Section IV. Finally, we draw  our conclusions in Section V.

\emph{Notations:} Throughout the paper, matrices, vectors, and scalars are denoted by letters of bold upper case, bold lower case, and lower case, respectively. Signs $(\cdot)^T$, $(\cdot)^H$, $\mid\cdot\mid$ and $\parallel\cdot\parallel$ represent transpose, conjugate transpose, modulus and norm, respectively. $\textbf{I}_N$ denotes the $N\times N$ identity matrix.

\section{System Model}
\begin{figure}
  \centering
  \includegraphics[width=0.4\textwidth]{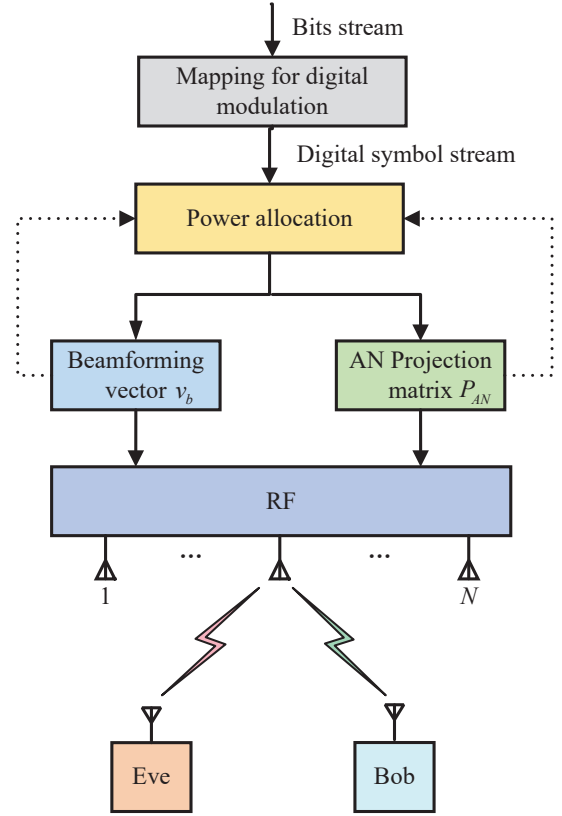}\\
  \caption{Schematic diagram of the proposed directional modulation system.}\label{Sys_Mod}
\end{figure}

The schematic diagram of the proposed DM system is illustrated in Fig.~\ref{Sys_Mod}, where Alice is equipped with $N$ antennas, and Bob and Eve are equipped with single antenna, respectively. In this paper, we assume there exists the line-of-sight (LOS) path. The transmitted baseband signal is expressed as
\begin{equation}\label{Tx signal s}
\mathbf{s}=\sqrt{\beta P_s}\mathbf{v}_bx+\sqrt{(1-\beta)P_s}\mathbf{P}_{AN}\mathbf{z},
\end{equation}
where $P_s$ is the total transmission power and limited, $\beta$ and  $(1-\beta)$ are the PA parameters of confidential messages and AN, respectively. $\mathbf{v}_b\in\mathbb{C}^{N\times1}$ denotes the transmit beamforming vector for controlling the confidential message to the desired direction and $\mathbf{P}_{AN}\in\mathbb{C}^{N\times N} $ is the projection matrix leading AN to the undesired direction, where $\mathbf{v}_b^H\mathbf{v}_b=1$ and $\mathrm{Tr}[\mathbf{P}_{AN}\mathbf{P}_{AN}^H]=1$. In (\ref{Tx signal s}),  $x$ is the confidential message of satisfying $\mathbb{E}\left\{x^Hx\right\}=1$ and $\mathbf{z}\in\mathbb{C}^{N\times1}$ denotes the AN vector with complex Gaussian distribution, i.e., $\mathbf{z}\sim\mathcal{C}\mathcal{N}(\mathbf{0},\mathbf{I}_{N})$.

Taking the path loss into consideration, the received signal at Bob can be written as
\begin{align}\label{Rx_signal yb}
y(\theta_b)
&=\sqrt{g_{ab}}\mathbf{h}^{H}(\theta_b)\mathbf{s}+n_b\nonumber\\
&=\sqrt{g_{ab}\beta P_s}\mathbf{h}^{H}(\theta_b)\mathbf{v}_bx\nonumber\\
&+\sqrt{g_{ab}(1-\beta)P_s}\mathbf{h}^{H}(\theta_b)\mathbf{P}_{AN}\mathbf{z}+n_b,
\end{align}
where $g_{ab}=\frac{\alpha}{d_{ab}^c}$ and $\mathbf{h}(\theta_b)\in\mathbb{C}^{N\times1}$ represent the path loss coefficient and channel vector between Alice and Bob, respectively. $d_{ab}$ is the distance between them, $c$ is the path loss exponent and $\alpha$ is the attenuation at reference distance $d_0$. $n_b$ is the complex additive white Gaussian noise (AWGN) with distribution $n_b\sim\mathcal{C}\mathcal{N}(0,\sigma_b^2)$.
Likewise, the received signal at Eve is given by
\begin{align}\label{Rx_signal ye}
y(\theta_e)
&=\sqrt{g_{ae}}\mathbf{h}^{H}(\theta_e)\mathbf{s}+n_e\nonumber\\
&=\sqrt{g_{ae}\beta P_s}\mathbf{h}^{H}(\theta_e)\mathbf{v}_bx\nonumber\\
&+\sqrt{g_{ae}(1-\beta)P_s}\mathbf{h}^{H}(\theta_e)\mathbf{P}_{AN}\mathbf{z}+n_e,
\end{align}
where $g_{ae}=\frac{\alpha}{d_{ae}^c}$, $d_{ae}$ and $\mathbf{h}(\theta_e)\in\mathbb{C}^{N\times1}$ denote the path loss coefficient, the distance and  the channel vector between Alice and Eve, respectively. $n_e$ is the complex AWGN following distribution $n_e\sim\mathcal{C}\mathcal{N}(0,\sigma_e^2)$. In the following, we assume that $\sigma_b^2=\sigma_e^2=\sigma^2$.

As per (\ref{Rx_signal yb}) and (\ref{Rx_signal ye}), we can derive the achievable rate along Bob and Eve as
\begin{equation}\label{Rb}
R(\theta_b)= \log_2\left(1+\frac{g_{ab}\beta P_s |\mathbf{h}^H(\theta_b)\mathbf{v}_b|^2}{g_{ab}(1-\beta) P_s \|\mathbf{h}^H(\theta_b)\mathbf{P}_{AN}\|_2^2+\sigma^2}\right)
\end{equation}
and
\begin{equation}\label{Re}
R(\theta_e)= \log_2\left(1+\frac{g_{ae}\beta P_s |\mathbf{h}^H(\theta_e)\mathbf{v}_b\|^2}{g_{ae}(1-\beta)|^2 P_s \|\mathbf{h}^H(\theta_e)\mathbf{P}_{AN}\|_2^2+\sigma^2}\right),
\end{equation}
respectively, which yield the following achievable secrecy rate (SR) $R_s$
\begin{equation}\label{Rs}
R_s=\max\left\{0,R(\theta_b)-R(\theta_e)\right\}.
\setcounter{equation}{6}
\end{equation}

\section{Proposed PA Strategy of Max-SR}
In this section, fixing $\mathbf{v}_b$ and $\mathbf{P}_{AN}$, the PA strategy of maximizing SR is proposed and its closed-form expression is presented. The derived expression is a general expression, which is suitable for  any beamforming scheme. For instance,  when the NSP beamforming scheme in \cite{Hu2016} is adopted, a simple formula  of OPA is directly given.

\subsection{Proposed General Power Allocation Strategy of Max-SR}
Before investigating PA, let us consider the joint optimization problem of Max-SR, which is casted as
\begin{align}\label{P1}
\mathrm{(P1):}&\max_{\mathbf{v}_b, \mathbf{P}_{AN}, \beta}~~~~R_s(\beta)=R(\theta_b)- R(\theta_e)\nonumber\\
&~~\text{s. t.}~~0\leqslant\beta\leqslant1\nonumber\\
&~~~~~~~~\mathbf{v}_b^H\mathbf{v}_b=1\nonumber\\
&~~~~~~~~\mathrm{Tr}[\mathbf{P}_{AN}\mathbf{P}_{AN}^H]=1.
\end{align}
where the three optimization variables are the PA factor  $\beta$, $\mathbf{v}_b$ and $\mathbf{P}_{AN}$. It is hard to solve the above optimization problem. In what follows, we focus on the PA problem by assuming that the beamforming scheme is given. For any fixed beamforming scheme, it is obvious that PA is an efficient and important way to enhance its SR.
In this subsection, we consider the design of power allocation factor $\beta$ based on maximizing the secrecy rate in general case. If the $\mathbf{v}_b$ and $\mathbf{P}_{AN}$ are known or designed well  in advance, then the above optimization degenerates towards the following simple PA problem.
\begin{align}\label{P2}
\mathrm{(P2):}&\max_{\beta}~~~~R_s(\beta)=R(\theta_b)- R(\theta_e)\nonumber\\
&~~\text{s. t.}~~0\leqslant\beta\leqslant1\nonumber\\
\end{align}
According to (\ref{Rb}) and (\ref{Re}), we can obtain the corresponding objective function $R_s(\beta)$ in (\ref{P2}) as
\begin{align}\label{Rs_beta}
R_s(\beta)
&=\log_2\frac{\overbrace{I\beta^2+J\beta+K}^{a(\beta)}}{\underbrace{L\beta^2+M\beta+K}_{b(\beta)}}\nonumber\\
&=\log_2\frac{a(\beta)}{b(\beta)}\nonumber\\
&=\log_2\phi(\beta),
\end{align}
where
\begin{align}
I&=g_{ab}g_{ae}P_s^2 \|\mathbf{h}^H(\theta_e)\mathbf{P}_{AN}\|_2^2\nonumber\\
&\times\left(\|\mathbf{h}^H(\theta_b)\mathbf{P}_{AN}\|_2^2-|\mathbf{h}^H(\theta_b)\mathbf{v}_b|^2 \right),\\
J&=g_{ab}P_s\left(|\mathbf{h}^H(\theta_b)\mathbf{v}_b|^2-\|\mathbf{h}^H(\theta_b)\mathbf{P}_{AN}\|_2^2\right)\nonumber\\
&\times\left(g_{ae}P_s \|\mathbf{h}^H(\theta_e)\mathbf{P}_{AN}\|_2^2+\sigma^2\right)\nonumber\\
&-g_{ae}P_s \|\mathbf{h}^H(\theta_e)\mathbf{P}_{AN}\|_2^2
\left(g_{ab}P_s \|\mathbf{h}^H(\theta_b)\mathbf{P}_{AN}\|_2^2+\sigma^2\right),\\
K&=\left(g_{ab}P_s \|\mathbf{h}^H(\theta_b)\mathbf{P}_{AN}\|_2^2+\sigma^2\right)\nonumber\\
&\times\left(g_{ae}P_s \|\mathbf{h}^H(\theta_e)\mathbf{P}_{AN}\|_2^2+\sigma^2\right),\\
L&=g_{ab}g_{ae}P_s^2 \|\mathbf{h}^H(\theta_b)\mathbf{P}_{AN}\|_2^2\nonumber\\
&\times\left(\|\mathbf{h}^H(\theta_e)\mathbf{P}_{AN}\|_2^2-|\mathbf{h}^H(\theta_e)\mathbf{v}_b|^2 \right),\\
M&=g_{ae}P_s\left(|\mathbf{h}^H(\theta_e)\mathbf{v}_b|^2-\|\mathbf{h}^H(\theta_e)\mathbf{P}_{AN}\|_2^2\right)\nonumber\\
&\times\left(g_{ab}P_s \|\mathbf{h}^H(\theta_b)\mathbf{P}_{AN}\|_2^2+\sigma^2\right)\nonumber\\
&-g_{ab}P_s \|\mathbf{h}^H(\theta_b)\mathbf{P}_{AN}\|_2^2
\left(g_{ae}P_s \|\mathbf{h}^H(\theta_e)\mathbf{P}_{AN}\|_2^2+\sigma^2\right).
\end{align}
Under the total  transmit power constraint, the SR given by (\ref{Rs_beta}) is also limited. This means
the numerator $a(\beta)$ and denominator $b(\beta)$ of the fraction inside logarithm operation (\ref{Rs_beta}) should be not equal to zero. Otherwise, an infinite value of SR is generated. Maximizing SR in (\ref{Rs_beta}) is equivalent to
\begin{align}\label{Rs_beta_deri}
\frac{\partial R_s(\beta)}{\partial\beta}=\frac{1}{\phi(\beta)}\frac{\partial\phi(\beta)}{\partial\beta}=0,
\end{align}
which can be reduced to
\begin{equation}\label{first}
\frac{\partial\phi(\beta)}{\partial\beta}=\frac{(IM-JL)\beta^2+2K(I-L)\beta+K(J-M)}{(L\beta^2+M\beta+K)^2}=0
\end{equation}
considering $\phi(\beta)\ne0$, where $\phi(\beta)=0$ means that SR is infinity. In terms of the above identity, we have the candidates for the optimal PA factor
\begin{align}\label{beta1}
\beta_1=\frac{-K(I-L)+\sqrt{\Delta}}{(IM-JL)}
\end{align}
and
\begin{align}\label{beta_2}
\beta_2=\frac{-K(I-L)-\sqrt{\Delta}}{(IM-JL)},
\end{align}
where $\Delta=K^2(I-L)^2)-K(IM-JL)(J-M)\geq0$ due to the non-negative real PA factor. The condition $a(\beta)=0$  yields two singular points
\begin{align}\label{beta_a}
\beta_{a1}=\frac{-J\pm\sqrt{J^2-4IK}}{2I}, \beta_{a2}=\frac{-J\pm\sqrt{J^2-4IK}}{2I}
\end{align}
The condition $b(\beta)=0$  yields the remain two singular points
\begin{align}\label{beta_b}
\beta_{a1}=\frac{-J\pm\sqrt{J^2-4IK}}{2I}, \beta_{a2}=\frac{-J\pm\sqrt{J^2-4IK}}{2I}
\end{align}
The above four critical points make the value of SR approach infinity, which is impossible to achieve an infinite SR with finite power in practice. For the purpose of simplifying our analysis below, it is assumed that the above four critical points lie outside the PA interval $[0,~1]$.
%

Meanwhile, we need to judge whether the two stationary points are in the interval of $(0,1)$. After that, we can obtain the optimal value of $\beta$ by comparing the values of $\phi(\beta)$ at endpoints and corresponding stationary points. While $\Delta\geq0$ cannot be guaranteed, we need to discuss the relation between 0 and $(IM-JL)$. The OPA parameter $\beta^*$ can be obtained by evaluating the following three cases.\\
\textbf{Case 1.} If $IM-JL>0$, $\phi(\beta)$ is a monotonously increasing function. Therefore, the OPA parameter is $\beta^*=1$ and the maximum secrecy rate is $R_s^*=R_s(1)=log_2\frac{I+J+K}{L+M+K}$, i.e., all power of Alice is employed to transmit confidential information and the AN fails to work.\\
\textbf{Case 2.} When $IM-JL=0$, the stationary point is $\beta_3=\frac{M-J}{2(I-L)}$. If $\beta_3\in(0,1)$, We need to compare the value of $\phi(0)$, $\phi(\beta_3)$ and $\phi(1)$ and obtain the OPA parameter $\beta^*$ by the corresponding value of $\beta$ of the maximum $\phi(\beta)$. Otherwise, we just need to compare the values of $\phi(0)$ and $\phi(1)$.\\
\textbf{Case 3.} If $IM-JL<0$, $\phi(\beta)$ is a monotonously decreasing function. Consequently, the OPA parameter is $\beta^*=0$ and the optimal secrecy rate is $R_s^*=R_s(0)=0$, i.e., no confidential messages is transmitted to Bob and the secure communication cannot be guaranteed.

Furthermore, the detailed operational procedures of the  proposed Max-SR PA strategy are presented in Algorithm~1.
\begin{algorithm}
\begin{enumerate}
  \item Initialization: $P_s$, $g_{ab}$, $g_{ae}$, $\mathbf{h}(\theta_b)$, $\mathbf{h}(\theta_e)$, SNR.
  \item Compute $\Delta=K^2(I-L)^2-K(IM-JL)(J-M)$.
  \begin{itemize}
    \item [$\mathrm{a)}$] If $\Delta\geq0$, four different cases are considered as fowllows\\
    \textbf{Case 1.} If $\beta_1\in(0,1)$, $\beta_2\in(0,1)$, then compare the values of $\phi(0)$, $\phi(\beta_1)$, $\phi(\beta_2)$ and $\phi(1)$.\\
    \textbf{Case 2.} If $\beta_1\in(0,1)$, $\beta_2\notin(0,1)$, then compare the values of $\phi(0)$, $\phi(\beta_1)$ and $\phi(1)$.\\
    \textbf{Case 3.} If $\beta_1\notin(0,1)$, $\beta_2\in(0,1)$, then compare the values of $\phi(0)$, $\phi(\beta_2)$ and $\phi(1)$.\\
    \textbf{Case 4.} If $\beta_1\notin(0,1)$, $\beta_2\notin(0,1)$, then compare the values of $\phi(0)$ and $\phi(1)$.\\
    After comparing the values of corresponding $\phi(\beta)$, we can get the OPA parameter $\beta^*$ by the corresponding value of $\beta$ of the maximum $\phi(\beta)$.
    \item [$\mathrm{b)}$] If $\Delta<0$, the OPA parameter $\beta^*$ has been solved in the aforementioned Case 1 to Case 3.
  \end{itemize}
\end{enumerate}
\caption{Proposed optimal power allocation strategy}\label{algorithm 1}
\end{algorithm}

\subsection{Proposed Simple Max-SR PA Strategy for NSP Beamforming Scheme}
If the general beamforming scheme in subsection A is the simplest NSP scheme, then the problem to compute the optimal PA factor $\beta$ can be significantly simplified. In the case of NSP,  the normalized values of beamforming vector $\mathbf{v}_b$ and projection matrix $\mathbf{P}_{AN}$ is given by \cite{Hu2016}
\begin{equation}\label{v_b}
 \mathbf{v}_b=\frac{1}{\sqrt{N}}\mathbf{h}(\theta_b)
\end{equation}
and
\begin{equation}\label{P_AN}
\mathbf{P}_{AN}=\frac{\mathbf{I}_N-\frac{1}{N}\mathbf{h}(\theta_b)\mathbf{h}^H(\theta_b)}{\|\mathbf{I}_N-\frac{1}{N}\mathbf{h}(\theta_b)\mathbf{h}^H(\theta_b)\|_F},
\end{equation}
respectively, where
\begin{equation}\label{h_theta_b}
\mathbf{h}(\theta_b)=\left[e^{j2\pi\Psi_{\theta_b}(1)}, \cdots, e^{j2\pi\Psi_{\theta_b}(n)}, \cdots, e^{j2\pi\Psi_{\theta_b}(N)}\right]^T
\end{equation}
and the phase function $\Psi_{\theta_b}(n)$ is defined as
\begin{equation}\label{var_phi}
\Psi_{\theta_b}(n)\triangleq-\frac{(n-(N+1)/2)d\cos\theta_b}{\lambda}, n=1,2,\cdots N,
\end{equation}
where $n$ denotes the $n$-th antenna, $d$ is the distance of two adjacent antennas, and $\lambda$ is the wavelength.

Substituting (\ref{v_b}) and (\ref{P_AN}) into (\ref{Rs_beta}), the first derivative of $\phi(\beta)$ can be written as
\begin{equation}\label{NSP_first}
\frac{\partial\phi(\beta)}{\partial\beta}=\frac{IM\beta^2+2KI\beta+K(J-M)}{(M\beta+K)^2},
\end{equation}
where
\begin{align}
I&=-Ng_{ab}g_{ae}P_s^2 \|\mathbf{h}^H(\theta_e)\mathbf{P}_{AN}\|_2^2,\\
J&=Ng_{ab}P_s\left(g_{ae}P_s \|\mathbf{h}^H(\theta_e)\mathbf{P}_{AN}\|_2^2+\sigma^2\right)\nonumber\\
&-g_{ae}P_s\sigma^2 \|\mathbf{h}^H(\theta_e)\mathbf{P}_{AN}\|_2^2,\\
K&=\sigma^2\left(g_{ae}P_s \|\mathbf{h}^H(\theta_e)\mathbf{P}_{AN}\|_2^2+\sigma^2\right),\\
M&=g_{ae}P_s\sigma^2\left(|\mathbf{h}^H(\theta_e)\mathbf{v}_b|^2-\|\mathbf{h}^H(\theta_e)\mathbf{P}_{AN}\|_2^2\right).
\end{align}
In the case of $IM\ne 0$, the corresponding roots $\beta_1$ and $\beta_2$  corresponding to the equation that  (25) is equal to zero can be denoted as
\begin{align}\label{NSP_beta1}
\beta_1=\frac{-KI+\sqrt{K^2 I^2-KIM(J-M)}}{IM}
\end{align}
and
\begin{align}\label{NSP_beta2}
\beta_2=\frac{-KI-\sqrt{K^2 I^2-KIM(J-M)}}{IM}
\end{align}
when $K^2 I^2-KI(J-M)\geq0$. In the case of $IM=0$, the stationary point $\beta_3$ can be formulated as
\begin{equation}\label{NSP_beta3}
\beta_3=\frac{M-J}{2I}.
\end{equation}
Based on the above results,  the determinant of OPA parameter $\beta^*$ is detailedly shown in Algorithm~1.

\section{Simulation and Discussion}
To assess the SR performance gain of the proposed Max-SR PA strategy, simulation results and analysis are presented in the following. Taking NSP as beamforming scheme, we numerically examine the effect of $\beta$ on the performance gain achieved by the optimal PA strategy in comparison with some typical PA strategies.

In our simulation, system parameters are set as follows: quadrature phase shift keying(QPSK) modulation, the total transmitting power $P_s=70$dBm, the spacing between two adjacent antennas $d=\lambda/2$,  the distance between Alice and Bob is $d_{ab}=500m$, the distance between Alice and Eve is $d_{ae}=500m$, the path loss exponent $c=2$, the desired direction $\theta_b=30^{\circ}$, and the eavesdropping direction $\theta_e=45^{\circ}$.

\begin{figure}
  \centering
  \includegraphics[width=0.5\textwidth]{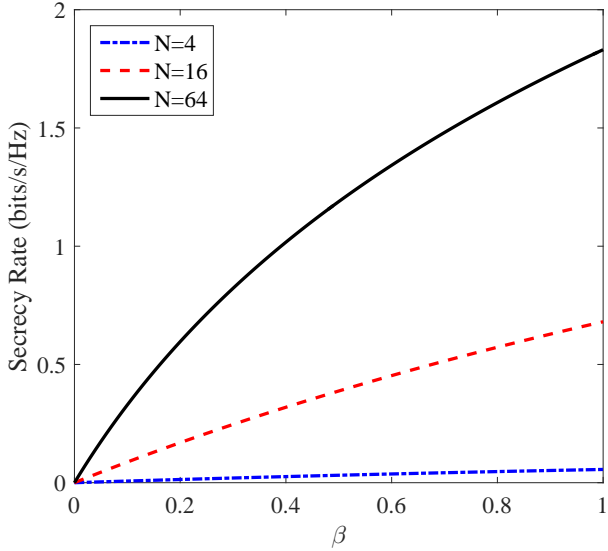}\\
  \caption{Secrecy rate versus $\beta$ (SNR=0dB).}\label{NSP_SR_beta_0dB}
\end{figure}

\begin{figure}
  \centering
  \includegraphics[width=0.5\textwidth]{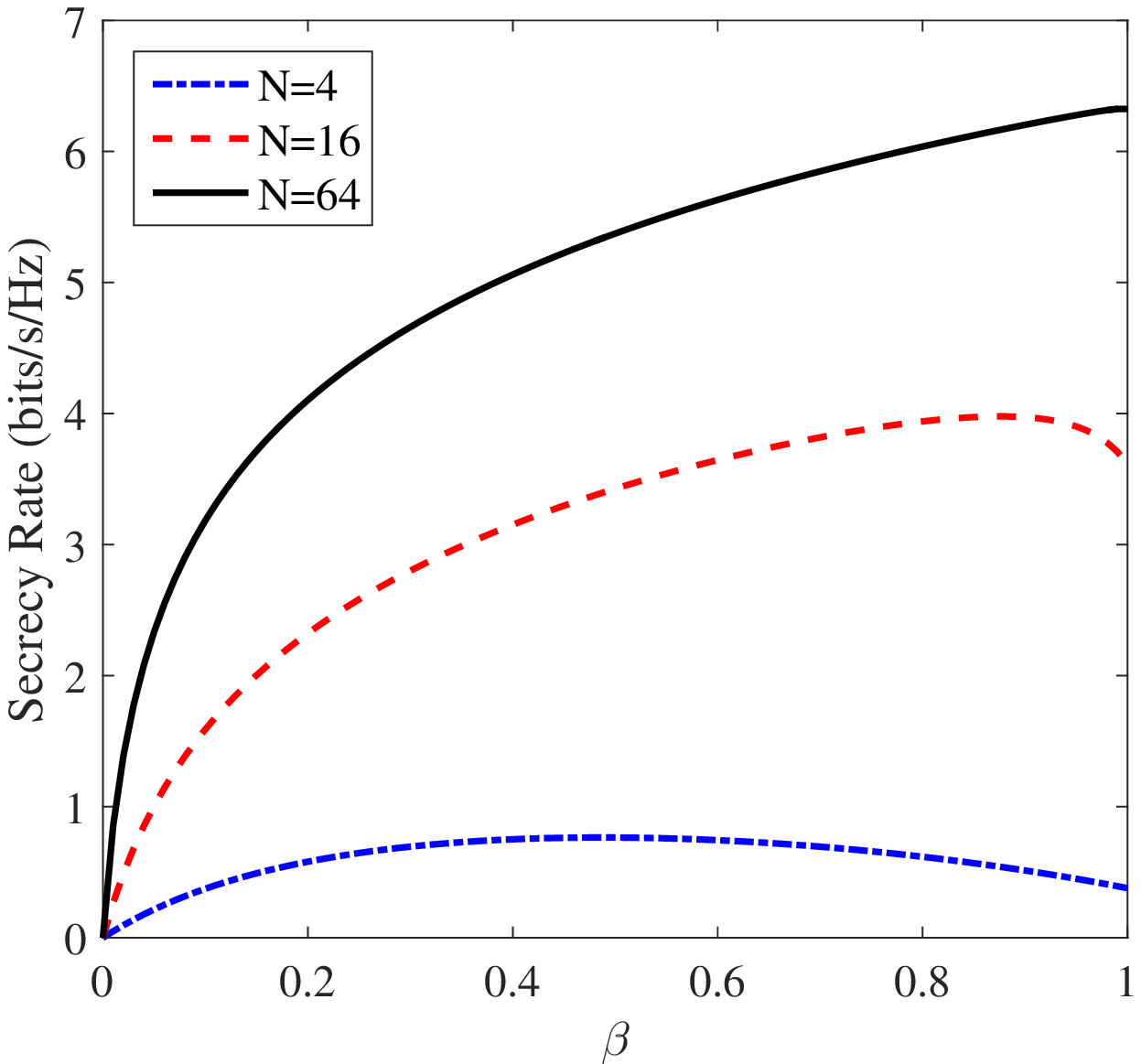}\\
  \caption{Secrecy rate versus $\beta$ (SNR=15dB).}\label{NSP_SR_beta_15dB}
\end{figure}

\begin{figure}
  \centering
  \includegraphics[width=0.5\textwidth]{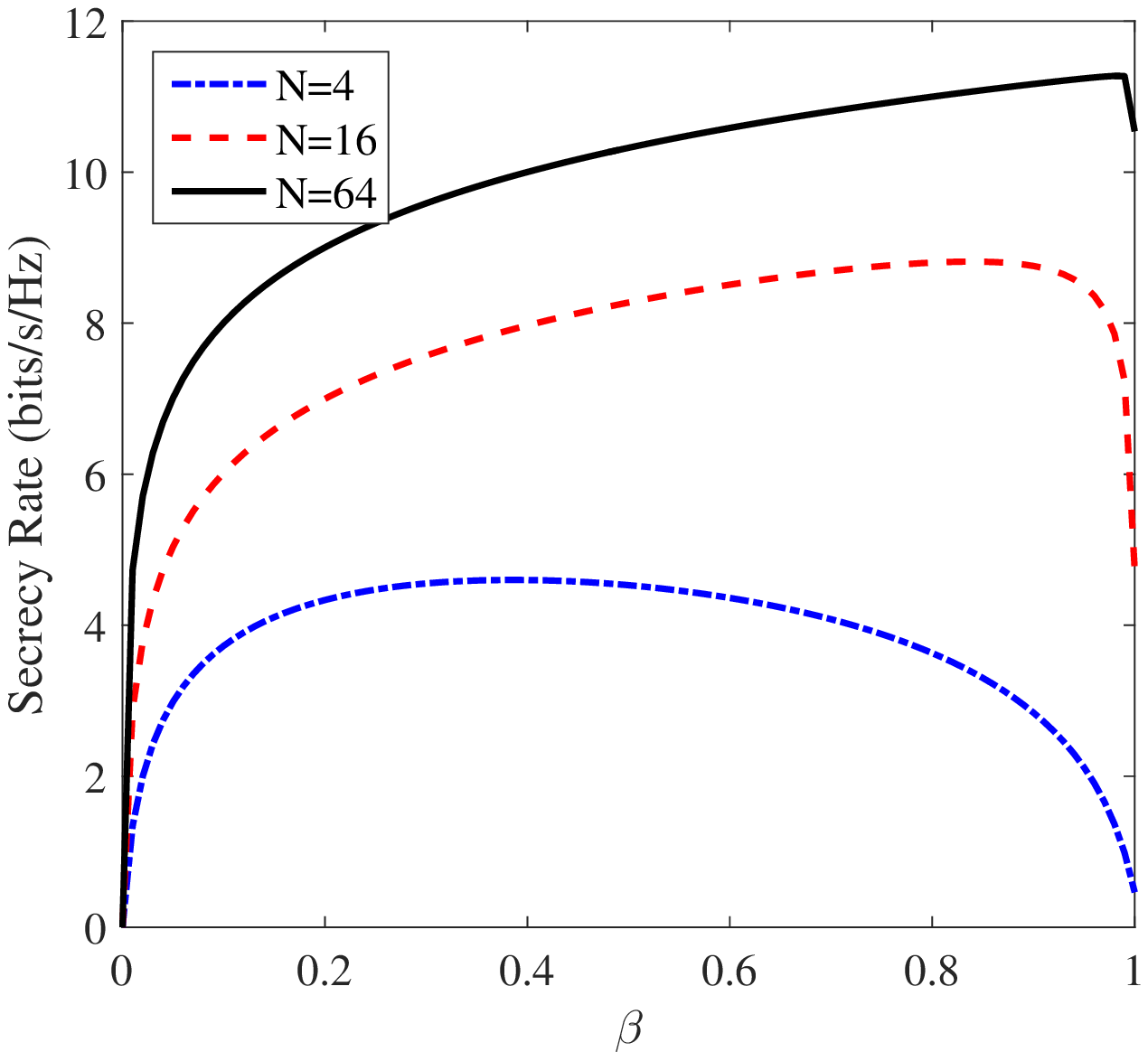}\\
  \caption{Secrecy rate versus $\beta$ (SNR=30dB).}\label{NSP_SR_beta_30dB}
\end{figure}


\begin{figure}
  \centering
  \includegraphics[width=0.5\textwidth]{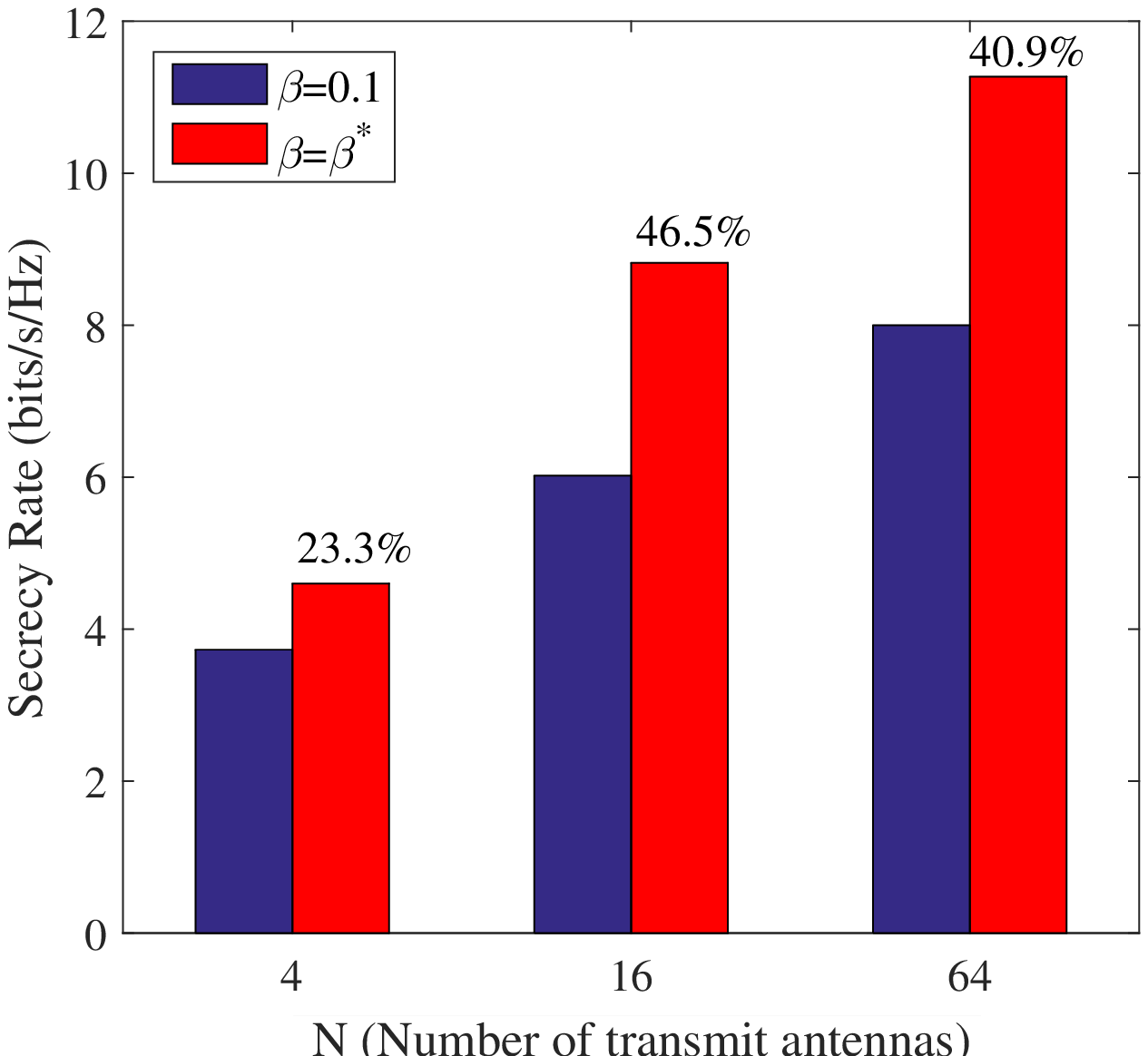}\\
  \caption{Secrecy rate versus N at $\beta$=0.1 and SNR=30dB}\label{NSP_SR_N_0.1}
\end{figure}

\begin{figure}
  \centering
  \includegraphics[width=0.5\textwidth]{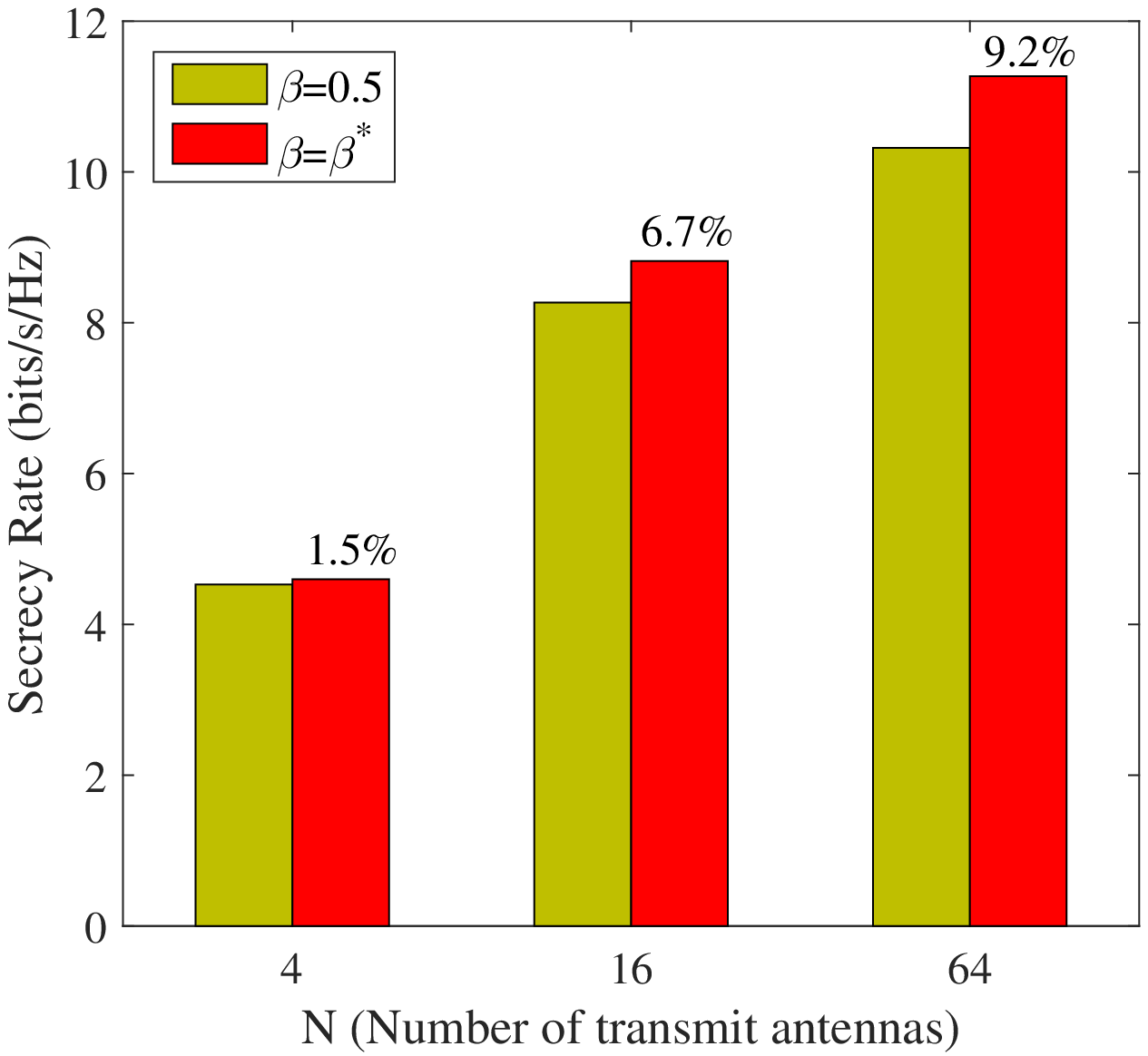}\\
  \caption{Secrecy rate versus N at $\beta$=0.5 and SNR=30dB}\label{NSP_SR_N_0.5}
\end{figure}

\begin{figure}
  \centering
  \includegraphics[width=0.5\textwidth]{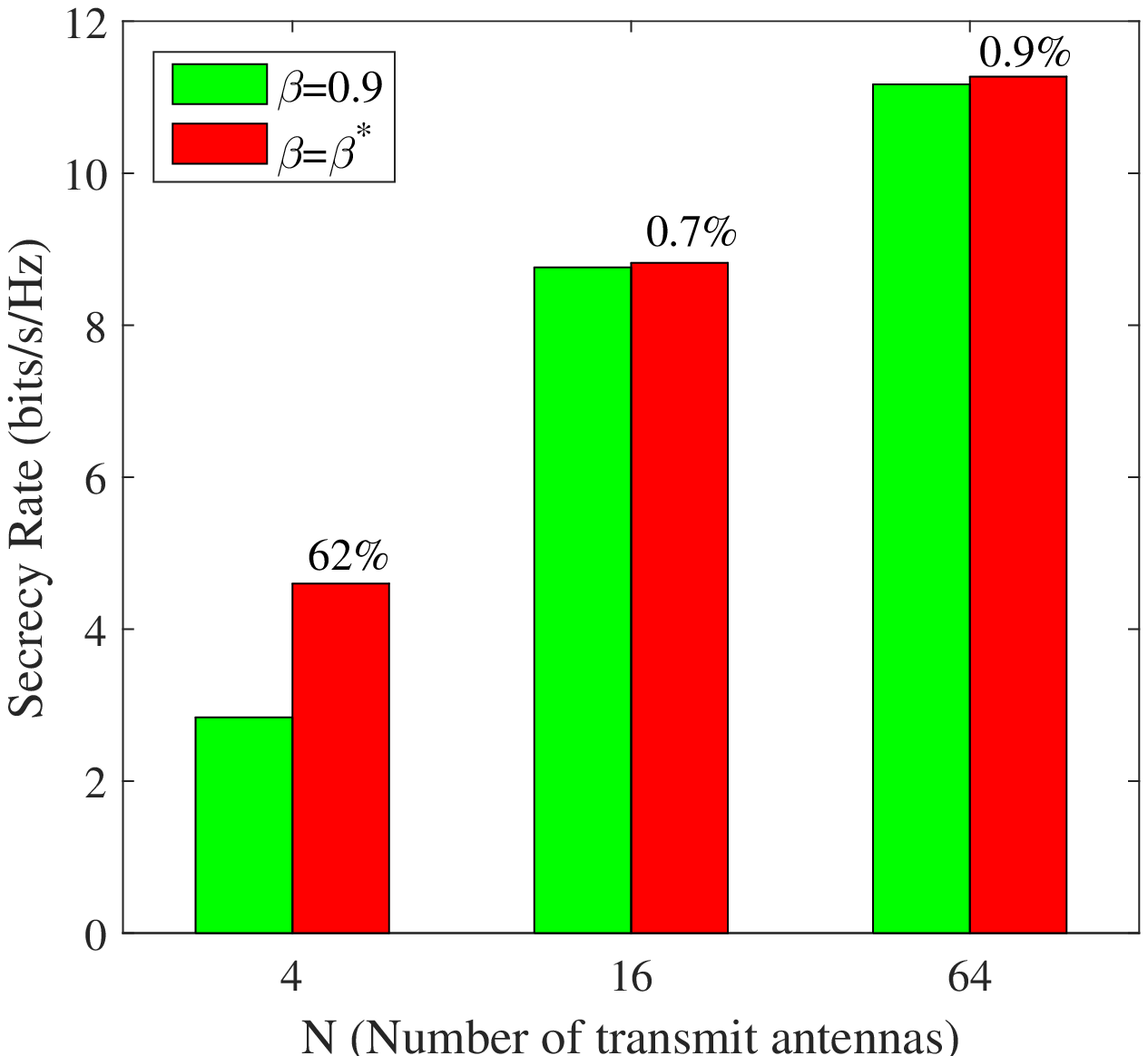}\\
  \caption{Secrecy rate versus N at $\beta$=0.9 and SNR=30dB}\label{NSP_SR_N_0.9}
\end{figure}

\begin{figure}
  \centering
  \includegraphics[width=0.5\textwidth]{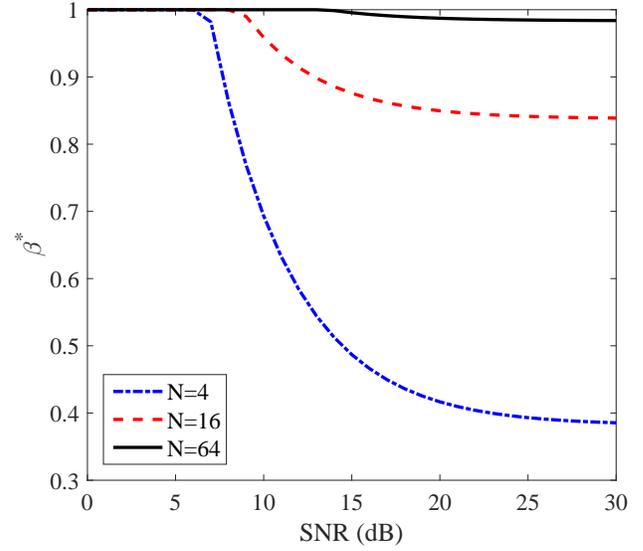}\\
  \caption{Theoretical optimal power allocation parameter $\beta^*$ versus SNR with different numbers of antennas.}\label{NSP_beta_SNR}
\end{figure}

Fig.~\ref{NSP_SR_beta_0dB} plots the curves of  SR (i.e., secrecy rate) versus $\beta$ with SNR=0dB. This case corresponds to the low SNR region.  From this figure,  we can observe that the SR increases with  increasing the number of antennas. In addition,  from the figure, it is seen that the SR increases continuously with the increase of $\beta$ regardless of the number of antennas. The result could be explained as follows: the superimposed AN is unnecessary in the low SNR regime because the channel noise is large enough. When the number of antennas is small, the SR is also small. For example, the SR approaches zero when the number of antennas is 4.

Fig.~\ref{NSP_SR_beta_15dB} demonstrates the curves of SR versus $\beta$ with SNR=15dB. The situation corresponds to the medium SNR region.  We can find that the SR curve is concave downward and a concave  function of $\beta$, i.e., there exists a unique value of OPA parameter $\beta^*$, which may maximize over  the interval [0,1]. Furthermore, the associated optimal value of $\beta$ are 0.49, 0.88 and 0.99 when the number of antennas is 4, 16 and 64. It is seen that the optimal value of $\beta$ grow gradually as the number of antennas increases from 4 to 64.

Fig.~\ref{NSP_SR_N_0.1}, Fig.~\ref{NSP_SR_N_0.5}, and Fig.~\ref{NSP_SR_N_0.9} depict the histograms of the SR versus N at $\beta=0.1$, $\beta=0.5$ and $\beta=0.9$, respectively. From Fig.~\ref{NSP_SR_N_0.1}, we can observe that the SR performance improvements achieved by the optimal value  $\beta$ over $\beta=0.1$  are  23.3\%, 46.5\% and 40.9\% for N=4, N=16, and N=64, respectively. The achievable SR performance gain is very attractive. Increasing the value of  $\beta$  up to $0.5$ even $0.9$, the SR performance gain indicates an obvious reduction trend as shown in Fig.~\ref{NSP_SR_N_0.5}, and Fig.~\ref{NSP_SR_N_0.9}. However, there are still a substantial performance gain in a small-scale number of transmit antennas.

Furthermore, the SR performance gain percentage compared with the SR of OPA parameter is shown in the three bar diagrams. The performance gain percentage achieved by OPA parameter at SNR=30dB is remarkable, especially when the PA factor $\beta=0.1$.

In Fig.~\ref{NSP_beta_SNR}, we plot the theoretical optimal power allocation parameter $\beta^*$ versus SNR with different numbers of antennas. Observing this figure, we can find that the theoretical OPA parameter $\beta^*$ maintain 1 in the low SNR regime and begin to decrease in the medium and high SNR regimes, which represents that the AN has a little effect on SR performance in the low SNR regime and the impact of AN becomes larger in the medium and high SNR regimes. Additionally, the theoretical OPA parameter $\beta^*$ declines earlier and faster for a small number of antennas, for example, $N=4$, compared with a larger number of antennas such as $N=16$ or $64$.  This demonstrates that the power to transmit confidential messages using a small number of antennas is less than that using a large number of antennas.

\section{Conclusion}
In our work, we proposed an optimum PA strategy of maximizing SR  in secure DM  networks. Firstly, a general optimization problem of maximizing SR was established. Given any beamforming scheme, the closed-form OPA strategy is given.  Then the OPA parameter can be obtained by discussing different scenarios. Finally, the NSP-based OPA strategy is taken into consideration and its closed-form formula was derived. Simulation and numerical results show that, in medium and high SNR regions, the proposed OPA can substantially improve the SR performance compared with some typical PA factors such as 0.1, 0.5, and 0.9.  Moreover, the OPA factor achieves its optimal value in the open interval $(0,~1)$,  and grows gradually with increasing in the number of transmit antennas. In the low SNR region,  the OPA factor $\beta$ is equal to 1, i.e., the total transmit power  is utilized to transmit confidential messages and the AN play a trivial role here. As the number of antennas tends to large-scale, the OPA parameter is close to one. In summary,  in the medium and high SNR regions or small number of transmit antennas at DM transmitter, the OPA strategy has an important impact on SR performance. The proposed OPA strategy is a general-form PA strategy suitable for any beamforming scheme, and  may be applied to any given bemaforming scheme. This makes it available to the diverse future applications  such as future mobile communications, satellite communications, millimeter-wave communications, unmanned-aerial-vehicles networks, device-to-device, and vehicle-to-vehicle.

\ifCLASSOPTIONcaptionsoff
  \newpage
\fi

\bibliographystyle{IEEEtran}
\bibliography{IEEEfull,cite}

\end{document}